\documentclass{aa}
\usepackage{graphicx}
\usepackage{amsmath}   
\usepackage{amssymb}
\usepackage{times}

\usepackage{color}

\def\eq#1{{Eq.~(\ref{#1})}}

\begin{document}

\title{Intensity mapping of Loeb-Rybicki haloes from scattering of galactic Lyman-$\alpha$ emission by the diffuse intergalactic medium before reionization}
\titlerunning{Intensity mapping of Lyman-$\alpha$ haloes before reionization}

 \author{Hamsa Padmanabhan
          \inst{1,2}
          \and
          Abraham Loeb\inst{3}
          }

   \institute{D\'epartement de Physique Th\'eorique, Universit\'e de Gen\`eve \\
24 quai Ernest-Ansermet, CH 1211 Gen\`eve 4, Switzerland
\and 
 The Green Concept - Institute for Carbon Assessments and Restoration Ecology,  India\\
              \email{hamsa.thegreenconcept@zohomail.in}
         \and
            Astronomy department, Harvard University \\
60 Garden Street, Cambridge, MA 02138, USA \\
             \email{aloeb@cfa.harvard.edu}            
             }

   \date{}


\abstract
 {
 We use the inferred evolution of Lyman-$\alpha$ luminosities of galaxies in the redshift range of  $z \sim 9-16$ from the \textit{James Webb Space Telescope}  (JWST) data to predict the power spectrum of Loeb-Rybicki haloes formed by { the scattering of Lyman-$\alpha$ photons from neutral hydrogen gas in the intergalactic medium expanding with the Hubble flow,  until they Doppler shift out of resonance and escape towards the observer.  This leads to the formation of the so-called Loeb-Rybicki intergalactic haloes, which are expected to be prominent even before the epoch of reionization.} We find excellent prospects for a statistical detection of the intensity mapping signal { from the clustering of these haloes,} with current and future experiments such as the SPHEREx and CDIM.  We also describe the detectability of the signal in cross-correlation with the 21-cm emission from {the neutral  hydrogen} in the intergalactic medium at these redshifts. We find that the cross-correlation signal should be detectable at a significance of a few { to a few tens} of standard deviations { out to $z \sim 13$ and marginally out to $z \sim 16$}, using the above experiments in combination with the Square Kilometre Array (SKA)-LOW  and its pathfinder, the Murchison Widefield Array (MWA).  
 }


\maketitle

\section{Introduction}

Intensity mapping,  in which the large-scale structure of the Universe is probed via the integrated emission of a spectral line -- without individually resolving luminous sources -- has emerged as a promising observational technique in the last decade \citep[for a recent review, see, e.g.,][]{kovetz2019}.  
By now, several theoretical and observational studies have focused on mapping spectral lines in the interstellar and intergalactic medium (IGM), notably the redshifted 21-cm line from atomic hydrogen (HI), an important tracer of physical processes in the early Universe and the formation of the first galaxies \citep{loeb2013}.  Mapping the evolution of HI is particularly important at and prior to the epoch of reionization, the second major phase transition of the Universe's baryons.  Observations indicate reionization to be nearly complete by redshifts 5.5-5.7 \citep[e.g.,][]{bosman2022} with its midpoint occurring around $z \sim 7.7$ \citep{planck2020}.  Statistical measurements of the clustering of HI in the post-reionization Universe have recently been made both in  intensity mapping auto-correlation \citep{paul2023},  as well as cross-correlations of intensity maps with galaxies \citep[][]{chang10, masui13, switzer13, wolz2022, anderson2018, amiri2024}, the latter of which { is helpful to mitigate the} contamination by foregrounds and interloper lines.

In addition to the 21 cm transition,  atomic hydrogen is characterized by the Lyman-$\alpha$ line,  its strongest
resonant spectral line with a rest wavelength of 1216 \AA. The Lyman-$\alpha$ transition of HI is observable both in absorption in the post-reionization Universe against the spectra of bright quasars \citep[the Lyman-$\alpha$ forest, which acts as a probe of the underlying density field, e.g.,][]{gunnpeterson, weinberg1997, fan2006}, and in emission as Lyman-$\alpha$ emitters \citep[LAEs;][]{hu1996, steidel1996} and diffuse Lyman-$\alpha$ blobs \citep[e.g.,][]{cantalupo2014, martin2014} in the intergalactic medium (IGM).  Intensity mapping of the Lyman-$\alpha$ line has been theoretically investigated for several decades \citep{partridge1967, gould1996, slosar2011, silva2013, wyithe2008, peterson2012, hutter2018, hutter2023, pullen2014} and recently measured in cross-correlation with other tracers of the IGM, such as quasars and the Lyman-$\alpha$ forest \citep{croft2016, croft2018} at lower redshifts,  viz. over $z \sim 0-3.5$. 

{ In the literature, there have been several models of Lyman-$\alpha$ emission arising from physical processes in the high-redshift IGM \citep{silva2013, pullen2014, heneka2017}.  Examples include that from cooling radiation in the cores of galaxies, recombinations in the interstellar and intergalactic medium, scattering of Lyman-$n$ photons which are re-emitted as Lyman-$\alpha$ photons, and collisional excitations followed by radiative cooling and partial re-emission in the Lyman-$\alpha$ frequency.}

{ In this paper, we consider another important source of Lyman-$\alpha$ radiation from the IGM, which is relevant even before reionization,  namely the scattering of Lyman-$\alpha$ photons by intergalactic neutral hydrogen gas in an (assumed) homogeneous,  uniform medium expanding with the Hubble flow,  until they redshift out of resonance and escape freely towards the observer.}  Before and during the epoch of reionization,  { this scattering around LAEs} leads to the development of the so-called Loeb-Rybicki intergalactic haloes \citep{loeb1999},  which are { prominently} observable as arcminute scale features in the cosmologically expanding neutral IGM.  { } In \citet[][hereafter Paper I]{hploeblaehaloes2024}, it was shown that such haloes may be detectable with the \textit{James Webb Space Telescope} (JWST) up to the pre-reionization epoch ($z \sim 10-20$), in the regime where the Lyman-$\alpha$ emission from their host galaxies is suppressed by  strong attenuation by the IGM \citep[the Gunn-Peterson effect,  ][]{gunnpeterson}. 

In this paper, we use the recently measured ultraviolet (UV) luminosities of galaxies in the pre-reionization epoch ($z \sim 9-16$) { from the JWST data} to forecast the statistical detectability of the Loeb-Rybicki haloes with intensity mapping, both in autocorrelation and cross-correlation with the diffuse 21 cm radiation from the HI in the IGM.  { For the Lyman-$\alpha$ intensity mapping, we consider the upcoming Spectro-Photometer for the History of the Universe,
EoR, and Ices Explorer \citep[SPHEREx;][]{dore2014} facility and a Cosmic Dawn Intensity Mapper \citep[CDIM;][]{cdimreport2019}-like configuration}.  For the 21 cm experiments, we consider an improved version of the Murchison Widefield Array (MWA) and its successor, the Square Kilometre Array (SKA)-LOW. We find strong prospects for the detectability of both the auto- and cross-power spectra in intensity mapping with the above experiments out to $z \sim 16$. 

The paper is organized as follows.  In Sec. \ref{sec:formalism}, we describe the formalism for measuring the { integrated} power spectrum of the Lyman-$\alpha$ intensity from { the clustering of the} Loeb-Rybicki haloes, in terms of the luminosities of { the LAE galaxies contributing the Lyman-$\alpha$ photons} as well as the density of { neutral hydrogen} in the IGM prior to and during the epoch of reionization.  { We describe how this signal is different from the other sources of diffuse Lyman-$\alpha$ emission from  galaxies and the IGM.} { The luminosities of the LAE galaxies can, in turn, be related to their host dark matter halo masses, by using the halo abundance matching technique calibrated to the JWST measurements, as developed in Paper I.} We then describe (Sec. \ref{sec:detectability}) the prospects for  detectability of this signal, both in auto-correlation as well as in cross-correlation with the intensity of 21 cm radiation from HI in the diffuse IGM, the latter of which is measurable with current and upcoming radio facilities { such as the SKA and its precursors}. We discuss the implications { in Sec. \ref{sec:discussion}. Throughout the paper, we use a $\Lambda$CDM cosmology having the matter density parameter $\Omega_m = 0.27$, the dark energy density parameter $\Omega_{\Lambda} = 0.73$, the baryon density parameter $\Omega_b = 0.042$, and the present-day Hubble parameter of $H_0 = 100 \ h \ $ km/s with $h = 0.71$.

\section{Intensity mapping with Lyman-$\alpha$ emission from Loeb-Rybicki haloes}
\label{sec:formalism}

{ \subsection{Radiation from Loeb-Rybicki haloes}

In the literature,  several physical processes leading to Lyman-$\alpha$ emission from the Interstellar and the Intergalactic Medium (ISM and IGM) at high redshifts have been considered \citep{silva2013, heneka2017, pullen2014}. These may broadly be classified into  (i) the formation of Lyman-Alpha Blobs, which mainly arise from cooling radiation from gas assembling into the cores of galaxies,  with sub-dominant contributions from continuum emission via stellar, free–free, free–bound, and two-photon processes,  and (ii) diffuse Lyman-alpha emission, due to photons generated by the IGM and ISM gas.  Possible sources for the latter which have been analysed in the literature include (a) recombinations of the gas in the ionized medium surrounding the source,  (b) scattering of Lyman-$n$ photons within the ISM and IGM which are then re-emitted as Lyman-$\alpha$ photons,  and (c) collisional excitations  followed by radiative cooling and partial re-emission in the Lyman-$\alpha$ line (including cascades from higher electronic levels). \footnote{ For a summary of the processes leading to the above sources of emission, see also Sec. 11.6.2 and 11.6.3 of \citet{loeb2013}.}

In this section, we provide a brief introduction to a different source of Lyman-$\alpha$ radiation,  namely that emitted by the formation of the so-called Loeb-Rybicki intergalactic haloes.   { The highest-redshift galaxies are observed to be strong Ly-$\alpha$ emitters \citep[e.g.,][]{dey1998}, likely due to their low dust content.  Here, we consider the transfer of Ly-$\alpha$ radiation in a uniform, fully neutral IGM undergoing a pure Hubble expansion around a steady point source   (considered to be an LAE galaxy) following \citet{loeb1999}.  Newly created Ly-$\alpha$ photons experience extremely high optical depth, and early scattering events redistribute photon frequencies symmetrically around the line center due to the isotropic thermal velocity distribution of hydrogen atoms \citep{dijkstra2014}.  As photons diffuse away from resonance and the medium becomes less opaque, the asymmetric redshift bias imposed by the Hubble expansion becomes dominant,  and the photons in the damping wing are shifted towards the red side of the resonance by the Doppler effect during scattering events.  Eventually, the photons redshift out of resonance and travel freely to the observer, who sees a faint Lyman-$\alpha$ halo around the source.  Since this transition occurs while the IGM remains highly opaque, we focus on the effect of Hubble expansion when calculating the observed intensity distribution.  In terms of the comoving frequencies, the scattering process may therefore be treated as effectively coherent (elastic). } 
It can be shown in this situation  that the optical depth of Lyman-$\alpha$ radiation from the source equals unity when the frequency redshift from the resonance, denoted by $\nu_{*}$, is given by $\nu_* = \beta/\alpha$, where
\begin{equation}
\alpha = H (z) \nu_{\rm Lya}/c  \ ; \beta = \left(\frac{3 c^2 \Lambda^2}{32 \pi^3 \nu_{\rm Lya}^2}\right) n_{\rm HI} \, .
\label{alphabeta}
\end{equation}
with $\nu_{\rm Lya} = 1216$ \AA \  being the rest wavelength of the Lyman-$\alpha$ line,  $c$ being the speed of light, $H(z)$ the Hubble parameter at the source redshift $z$ and $\Lambda =6.25 \times 10^8 $ s$^{-1}$ being the decay rate of the 2$p$ to 1$s$ transition of neutral hydrogen.  The $n_{\rm HI}$ denotes the number density of neutral hydrogen in the surrounding IGM.  This  corresponds to a halo in real space with a proper size of $r_* = \beta/\alpha^2$. 

In a uniform, dust free surrounding medium,  $n_{\rm HI} \equiv \bar{n}_{\rm HI}$ is the cosmological mean neutral hydrogen density,  expressible in terms of the baryon number density parameter, $\Omega_b$ as $\bar{n}_{\rm HI} = \Omega_b \rho_{c,0} (1 - Y_{\rm He}) (1 + z)^3/m_p $,  in which  $Y_{\rm He} = 0.24$ is the helium fraction by mass, and $m_p$ is the proton mass.  In this situation, 
 { it can be shown that $\nu_*$ is expressible as:
\begin{equation}
\nu_*  = 5.6 \times 10^{12} \ h \ \Omega_b \ (1+z)^{3} \ [\Omega_m (1+z)^3 + \Omega_{\Lambda}]^{-1/2} \ {\rm Hz}
\end{equation}
 which,  for {early matter domination, that happens at} $z \gg 1$ in a flat $\Lambda$CDM universe, reduces to}:  
\begin{equation}
\nu_* = 8.85 \times 10^{12} {\rm Hz} \times \left(\frac{\Omega_b h}{0.05 \sqrt{\Omega_m}}\right) \left(\frac{1+z}{10} \right)^{3/2}
\end{equation}
that corresponds (for $\nu_* = 8.85 \times 10^{12} {\rm Hz}$) to a Doppler velocity of $v = c \nu_*/\nu_{\alpha}  \sim 1000$ km/s. The corresponding proper size of the halo is found to be given by:
\begin{equation}
r_* = 2.1 \times 10^{25} \ {\rm cm} \frac{(\Omega_b/\Omega_m)}{1 + (\Omega_{\Lambda}/\Omega_m)(1+z)^{-3}}
\label{defrstar}
\end{equation}
This corresponds to an angular size of $\sim 15''$ on the sky for a halo at the source redshift $z \sim 10$.

The  specific intensity (in units of photons cm$^{-2}$ s$^{-1}$ sr$^{-1}$ Hz$^{-1}$) of such a Lyman-$\alpha$ halo in a uniform, spherical medium undergoing  Hubble expansion is written as ${I}(\tilde{p},\tilde{\nu})$ in terms of the dimensionless impact parameter $\tilde{p}$, defined as $\tilde{p} = \tilde{r} \sqrt{1-\mu^2}$, where $\mu \equiv \cos \theta$ is the direction of emission relative to the radius vector at radius $r$,  with $\tilde{r} \equiv r/r_*$ and the normalized frequency shift from the Lyman-$\alpha$ resonance being given by $\tilde{\nu} \equiv |\nu - \nu_{\rm Lya}|/\nu_*$.  
The intensity, in turn, is conventionally expressed\footnote{ Technically, we are calling $I$ here what was defined as $J$ in \citet{loeb1999}, hence the additional factor of $4 \pi$ in the denominator with respect to the expression in that work.} in units of $I_* = \dot{N}_{\alpha}/(4 \pi r_*^2)$, that is, $\tilde{I} = I/I_*$ where $\dot{N}_{\alpha}$ is the emissivity rate of Lyman-$\alpha$ photons. The intensity integrated over all frequencies, is, in turn, expressed as 
\begin{equation}
\tilde{I}(\tilde{p}) = \int d \tilde{\nu} \  \tilde{I}(\tilde{p}, \tilde{\nu})
\end{equation}
and measured in units of photons cm$^{-2}$ s$^{-1}$ sr$^{-1}$.
The above intensities are defined at the source frame.  { For conversion to the observer's frame,  $ \tilde{I}(\tilde{p}, \tilde{\nu})$ needs to be divided by $(1+z)^2$,  due to the fact that the phase space density, $I(\tilde{p},\tilde{\nu})/\nu^2$ is conserved during Hubble expansion.  Thus, $\tilde{I}(\tilde{p})$,  which contains one further factor of $\nu$},  needs to be divided by $(1+z)^3$.  A  Monte Carlo calculation can be used to derive the shape of the profile $\tilde{I}(\tilde{p})$ as a function of the dimensionless impact parameter, as shown in Fig. 1 of \citet{loeb1999}. The maximum of the intensity occurs at zero impact parameter, and its value is found to be $\tilde{I}(0) = 0.2$.

Putting it all together, the central intensity of the radiation from Loeb-Rybicki haloes, in units of ergs cm$^{-2}$ s$^{-1}$ sr$^{-1}$ is given by:
\begin{eqnarray}
 I_{\rm cen} &=&  \frac{\tilde{I}(0) h_{\rm P} \nu_{\rm obs}}{(1+ z)^3} \frac{\dot{N}_{\alpha}}{4 \pi r_*^2} \nonumber \\
 && {\rm erg s^{-1} cm^{-2} s^{-1} sr^{-1}} . 
\end{eqnarray} 
in which $\nu_{\rm obs} = \nu_{\rm Lya}/(1+z)$ is the observed frequency of the radiation in terms of the redshift of emission $z$ and $h_{\rm P}$ is Planck's constant.
The above now needs to be converted into the (observed) \textit{specific} intensity, i.e.  that measured in ergs cm$^{-2}$ s$^{-1}$ sr$^{-1} $ Hz$^{-1}$. For this, one typically considers the intensity spread over the observed bandwidth, or frequency interval.  The latter is given by the observed frequency width at half maximum of the Lyman-$\alpha$ line, which in this context is found to be $\Delta \nu = 0.5 \nu_*/(1+z)$ [see Paper I and \citet{loeb1999}].  This yields
\begin{eqnarray}
I_{\rm obs} =  \frac{I_{\rm cen}}{\Delta \nu} 
 &=& \frac{\tilde{I}(0) h_{\rm P} \nu_{\rm obs}}{(1+ z)^2} \frac{\dot{N}_{\alpha}(M, z)}{2 \pi r_*^2 \nu_*}  \nonumber \\
 && {\rm erg s^{-1} s^{-1} cm^{-2} Hz^{-1} sr^{-1}} . 
  \label{iobs}
\end{eqnarray}

The emissivity rate of Lyman-$\alpha$ photons,  $\dot{N}_{\alpha} (M,z) $,  is expressed above in terms of the host dark matter halo mass $M$ of the LAE galaxy, and its redshift $z$.  In the above context, it can be expressed as $\dot{N}_{\alpha}\equiv  L_{\rm Lya}(M,z)/h_{\rm P} \nu_{\rm Lya}$, in terms of the observed Lyman-$\alpha$ luminosity, $L_{\rm Lya}(M,z)$ of the galaxy.   The latter, in turn,  can be computed (as done in Paper I) by using abundance matching of the ultraviolet (UV) luminosity function at the redshift of interest,  as measured by the JWST NIRCam and NIRSpec data \citep{donnan2023, harikane2023} to the Sheth-Tormen dark matter halo mass function. The UV luminosities as a function of dark matter halo mass are used to infer the corresponding (intrinsic) Lyman-alpha luminosities, $L_{\rm Lya, int}(M,z)$ over $z \sim 9-16$. These are further scaled by a factor of $f_{\rm esc} \sim 0.15$, representing the mean  Lyman-$\alpha$ escape fraction at lower redshifts { \citep[e.g.,][]{hayes2011, lin2024, saxena2024}},   in order to obtain the observed{\footnote{ Of course,  the direct detection of Lyman-$\alpha$ from individual LAEs at these redshifts is strongly hampered by its attenuation in the IGM.  However,  the scattered radiation in the form of Loeb-Rybicki haloes was found to be detectable at the level of a few standard deviations with the JWST.}} Lyman-$\alpha$ luminosities as a function of halo mass and redshift: $L_{\rm Lya}(M,z) = f_{\rm esc} L_{\rm Lya, int}(M,z) $.  
}

{ \subsection{Intensity mapping of Lyman-$\alpha$ radiation from Loeb-Rybicki haloes}}

Substituting for the emissivity rate  of the Lyman-$\alpha$ emission $\dot{N}_{\alpha}= L_{\rm Lya}(M,z)/h_{\rm P} \nu_{\rm Lya}$  and using the central value of the intensity $\tilde{I}(0) = 0.2$ in \eq{iobs} yields:
\begin{eqnarray}
&& I_{\rm obs} ({M, z}) = \frac{0.2 {L}_{\rm Lya}(M,z)}{2 \pi r_*^2 \nu_* (1+ z)^3}  \, ,
\label{iobsnew}
\end{eqnarray}
where the units have been suppressed for brevity.
Given the definitions of $r_*$ and $\nu_*$,  this can be further simplified to read:
\begin{eqnarray}
I_{{\rm obs}}(M, z) &=& A \ {L}_{\rm Lya}(M,z) \ n_{\rm HI}^{-3}  \, ,
\end{eqnarray}
where $A$ is a function of fundamental constants and the source redshift $z$.

We now develop the formalism for analysing the fluctuations in $I_{\rm obs}$.  This is done by splitting the $n_{\rm HI}$ term into a mean and fluctuating component:
\begin{equation}
n_{\rm HI} = \bar{n}_{\rm HI} (1+ \delta_{\rm HI}) \ \,
\end{equation}
where $\delta_{\rm HI}  = b_{\rm HI} \delta_{\rm dm} \ll 1$ in the neutral hydrogen overdensity defined in terms of the linear large scale bias of neutral hydrogen, $b_{\rm HI}$.  In the above,  $\bar{n}_{\rm HI}(z) \equiv \Omega_b \rho_{c,0} (1 - Y_{\rm He}) (1 + z)^3/m_p$ is 
the average intergalactic HI density, { as defined in the previous subsection}.

We can similarly account for the fluctuations in the Lyman-$\alpha$ luminosity, denoted by $\delta L_{\rm Lya}$, to linearize the expression for $I_{\rm obs}$ as:
\begin{eqnarray}
&& I_{{\rm obs}}(M, z) \approx A \ {\bar{L}}_{\rm Lya}(M,z)   \ \bar{n}_{\rm HI}^{-3} \nonumber\\
&&\left(1 +  \delta L_{\rm Lya}\right) \left(1 - 3  \delta_{\rm HI}\right)  \nonumber \\
& \equiv & I_0 \left(1 +  \delta L_{\rm Lya}\right) \left(1 - 3  \delta_{\rm HI}\right) \, .
\end{eqnarray}

The (dimensionless) power spectrum of the above fluctuations, on scales greater than the halo size, can now be expressed in terms of the dark matter halo model (for the LAE emission) and that of a biased tracer of the dark matter (for the HI) as the sum of three terms (all implicitly functions of the redshift $z$):
\begin{equation}
P_{\rm HI-Lya,  2h} = \left(P_{\rm Lya} - 6 P_{\rm HI, Lya} + 9 P_{\rm HI} \right) \, ,
\label{twohalo}
\end{equation}
where the first two terms are related to the Lyman-$\alpha$ emission (defined below), and the last term denotes the power spectrum of the HI, given by
\begin{equation}
P_{\rm HI} =  P_{\rm m} (k) b_{\rm HI}^2 \, ,
\label{twohalohi}
\end{equation}
in which $P_{\rm m}(k)$ is the (dark) matter power spectrum,
and $b_{\rm HI}$ is the bias of the intergalactic HI defined above.

The first term in \eq{twohalo} denotes the power spectrum of the Lyman-$\alpha$ radiation coming from the LAE galaxies,  which can be connected to  host dark matter halo masses by the procedure described in Paper I.  This can be expressed as:
\begin{eqnarray}
P_{\rm Lya} &=& P_{\rm m} (k) \left[\frac{1}{\bar{L}_{\rm Lya}} \int_{M_{\rm min, Lya}}^{\infty} dM \  \frac{dn}{dM} \ L_{\rm Lya} (M) \ b (M) \right]^2 \nonumber \\
& \equiv &  b_{\rm LAE}^2  P_{\rm m} (k) \, ,
\label{Lyaspint}
\end{eqnarray} 
in which we have defined the large-scale bias of the Lyman-alpha emission from the LAEs in the last line (see also Paper I), and
\begin{equation}
\bar{L}_{\rm Lya} = \int_{M_{\rm min, Lya}}^{\infty} dM \  \frac{dn}{dM}  L_{\rm Lya} (M)/ \int_{M_{\rm min, Lya}}^{\infty} dM \  \frac{dn}{dM} 
\end{equation}

In both equations above,  $M_{\rm min, Lya}$ is the minimum halo mass probed at each redshift, which corresponds to the flux limit $f_{\rm lim}$ of the survey (see also Paper I; we assume $f_{\rm lim} = 10^{-17} {\rm erg/s/cm}^2$ for consistency with the clustering measurements described in that work). 
The second term in \eq{twohalo} denotes the cross-power spectrum between the HI density and the Lyman-$\alpha$ radiation, given by:
\begin{eqnarray}
P_{\rm HI, Lya} &=&  P_{\rm m} (k) b_{\rm LAE} b_{\rm HI}   \nonumber \\
\end{eqnarray}
where $b_{\rm LAE}$ is defined as in \eq{Lyaspint}.
On scales smaller than the size of a (dark matter) halo, the power spectrum is modulated by the (normalized) shot noise, coming from the LAE emission alone and which may be described by the $k \to 0$ limit  of the one-halo term \citep[e.g.,][]{hp2023hi}:
\begin{eqnarray}
P_{\rm Lya, 1h} &=&  \int_{M_{\rm min, Lya}}^{\infty} dM \  \frac{dn}{dM} \ L_{\rm Lya} (M)^2 \nonumber\\
&&  \left[\int_{M_{\rm min, Lya}}^{\infty} dM \  \frac{dn}{dM} \ L_{\rm Lya} (M)\right]^{-2} \, .
\label{lya1h}
\end{eqnarray}

\begin{figure}
    \centering
    \includegraphics[width =\columnwidth]{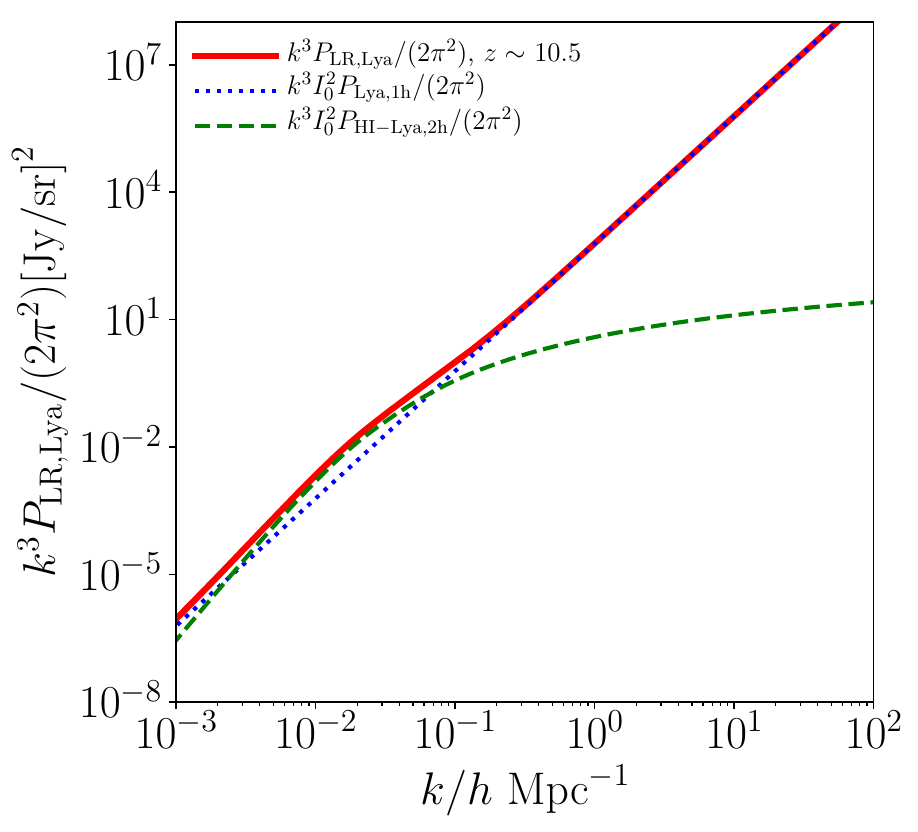}   \includegraphics[width =\columnwidth]{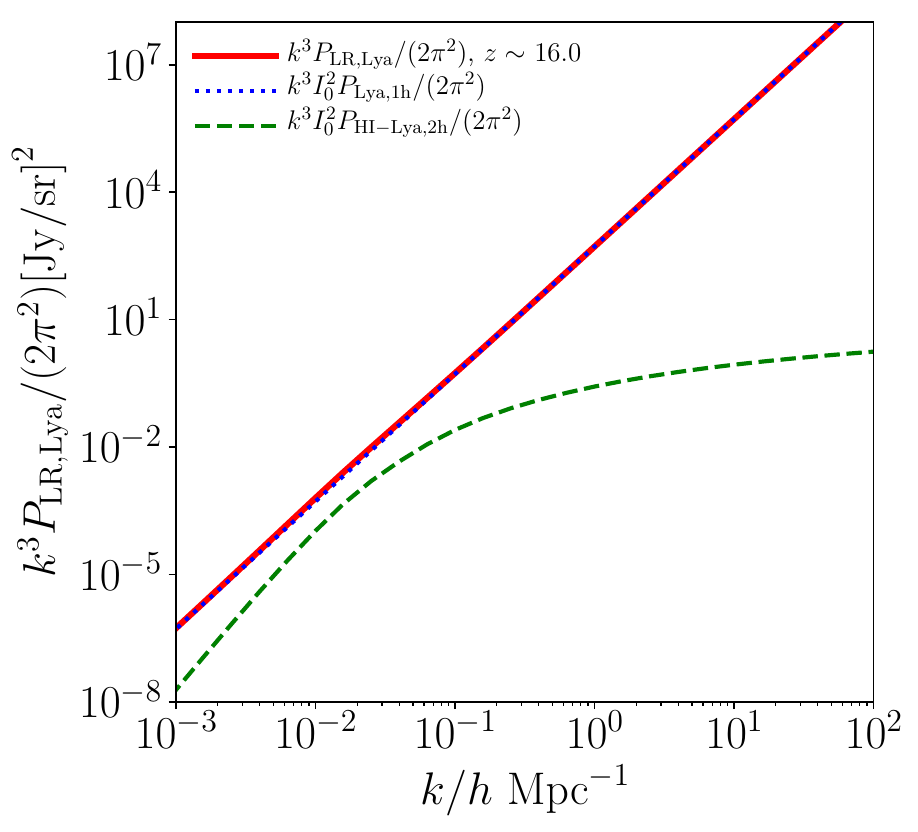}   
    \caption{Power spectrum  (thick red solid line) of the intensity from Loeb-Rybicki haloes, \eq{powerspeclr} at {two redshifts,  $z \sim 10.5$ (top panel) and $z \sim 16$ (lower panel)}. Overplotted are its component 1- and 2-halo terms (\eq{lya1h} and \eq{twohalo},  normalized by $I_0^2$), shown by the blue dotted and green dashed lines, respectively.}
   \label{fig:fidpower}
\end{figure}
Putting it all together, the full power spectrum of the scattered radiation in Loeb-Rybicki haloes can be written as:
\begin{equation}
 P_{\rm LR,Lya} =  I_0^2 \left[P_{\rm Lya, 1h} + P_{\rm HI-Lya,  2h} \right] \, .
\label{powerspeclr}
\end{equation} 
Note that the above expression is implicitly a function of both $k$ and $z$, as inherited from the dependencies of its component terms.   
Using the abundance matched relation between the Lyman-$\alpha$ luminosity and host dark matter halo mass derived from the JWST results (see Paper I),  and assuming a unit bias for intergalactic HI ($b_{\rm HI} = 1$), we can calculate the power spectrum in \eq{powerspeclr} at the redshifts of interest in the (pre)-reionization epoch, $z \sim 9, 10.5, 13, 16$.

The power spectrum thus calculated for a fiducial redshift $z \sim 10.5$ is plotted in {the top panel of Fig. \ref{fig:fidpower}.} Overplotted are the contributions of its component 1-halo and 2-halo terms,  $P_{\rm Lya, 1h}$ and $P_{\rm HI-Lya, 2h}$ defined above.  The `break' between the one- and two-halo terms occurs at a scale of a few Mpc, roughly corresponding to the characteristic size $r_*$ associated with Loeb-Rybicki haloes (see \eq{defrstar}.) 

{ A few comments regarding the redshift dependence of the signal are also in order.  From \eq{iobsnew} and the definitions of $r_*$ and $\nu_*$,  it can be shown that the combination of the $H(z)$ and $(1+z)$ terms leads to an redshift dependence of $(1+z)^{-4.5}$ in $I_0$. This variation with redshift,  of about an order of magnitude over the range $z \sim 9-16$,  is compensated in large part by the 1-halo term (\eq{lya1h}) sourced by $L_{\rm Lya}(M,z)$,  being the dominant part of the power spectrum (as seen from Fig. \ref{fig:fidpower}), which increases with increasing redshift by about 1.5 orders of magnitude on the relevant scales. {\footnote{The evolution of $L_{\rm Lya}(M,z)$  is obtained from the abundance matching of the inferred Lyman-$\alpha$ luminosity functions to their host dark matter halo masses, see, e.g.,  Fig. 1 of Paper I.}}  This overall effect is  a slow variation of the signal with redshift,  by about a factor of a few in the redshift range under consideration,  $z \sim 9$ to $z \sim 16$.} {This can be seen from the lower panel of Fig. \ref{fig:fidpower},  which shows the power spectrum at a different redshift, $z \sim 16$ along with its component terms.}

\section{Detectability of the signal}
\label{sec:detectability}

We now consider the detectability of the power spectrum of the scattered Lyman-$\alpha$ radiation by current and future experiments.  In Paper I, we had found that the {\it James Webb Space Telescope} would be able to detect the Lyman-$\alpha$ radiation from  individual (or stacked) Loeb-Rybicki haloes out to $z \sim 9-11$.  { We consider an upcoming ``Stage II" Lyman-$\alpha$ intensity mapping survey consistent with the Cosmic Dawn Intensity Mapper (CDIM)-like  or SPHEREx configurations.}

The noise of a  Lyman-$\alpha$ { intensity mapping} survey is given in terms of the instrument's pixel uncertainty,  $\sigma_{\rm N, Lya}$, typically expressed in ergs s$^{-1}$ cm$^{-2}$ sr$^{-1}$ Hz$^{-1}$ \citep[e.g.,][]{masribas2020}
and the pixel volume, $V_{\rm pix}$, as:
\begin{equation}
P^{\rm Lya}_{\rm N} = {\sigma_{\rm N, Lya}^{2}} V_{\rm pix} \; .
\label{noiselyaauto}
\end{equation}

\begin{table*}
\begin{center}
    \begin{tabular}{ | c | c | c | c | c |  p{3cm} |}
    \hline
    Configuration &  $D_{\rm dish}$ (m.)  & $\delta \nu$ (GHz)  & $S_{\rm A}$ (sq. deg.) & $\sigma_{\rm N}$ 
(erg cm$^{-2}$ s$^{-1}$ Hz$^{-1}$ / sr) & $B_{\nu}$   \\ \hline
CDIM-like &  0.83 & 400 &  300 & $3.16 \times 10^{-19}$  & 360000 GHz   \\ 
   SPHEREx &  0.20 & 1900 &  30000 & ${ 8.33 \times 10^{-20}}$  & 200000 GHz  \\ 
\hline
    \end{tabular}
\end{center}
\caption{ {Experimental parameters assumed for the { CDIM-like and  SPHEREx} survey designs. { Symbols have the meanings provided in the main text. }}}
 \label{table:improved}
\end{table*}

\begin{figure}
    \centering
    \includegraphics[width =\columnwidth]{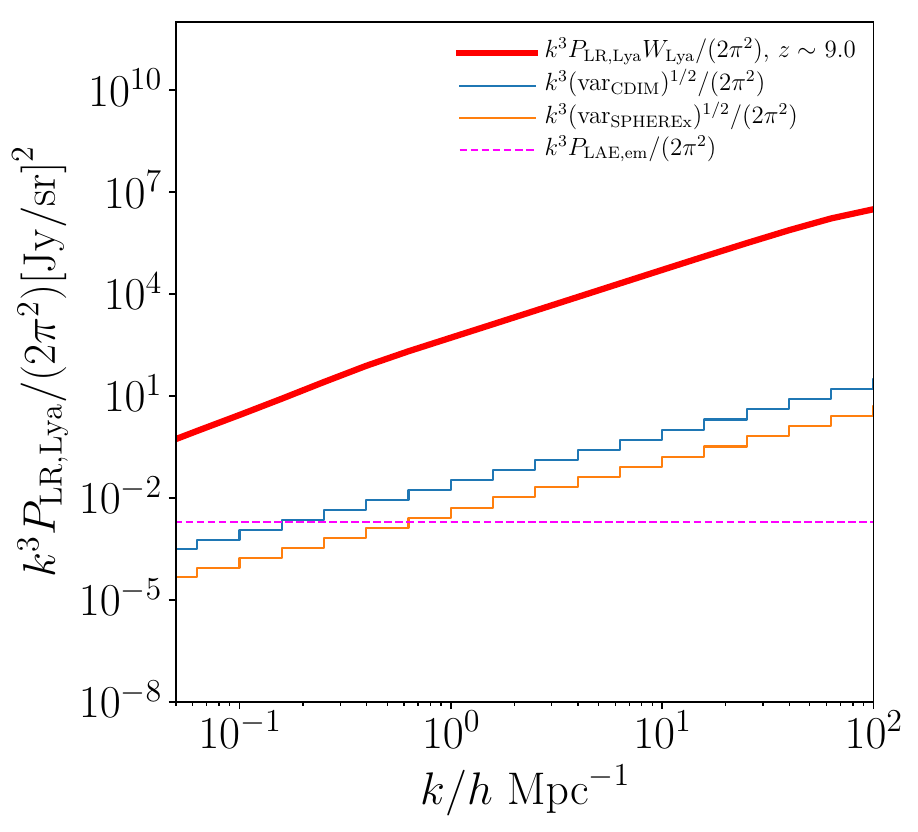}   \includegraphics[width = \columnwidth]{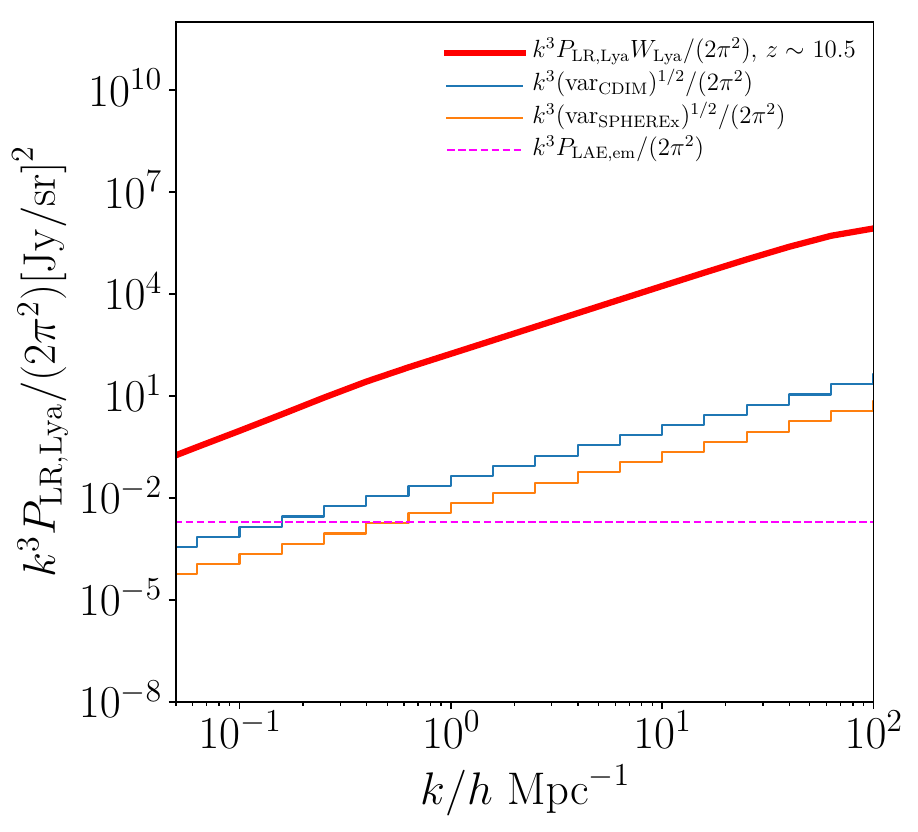} 
    \caption{ Signal and noise power in autocorrelation intensity mapping of Lyman-$\alpha$ haloes at $z \sim 9$ (top panel) and $z \sim 10.5$ (lower panel) probed by the CDIM-like or SPHEREx  surveys.  The square root of the noise variance, given by \eq{varianceauto},  is indicated by the thinner blue and orange steps for the CDIM-like and SPHEREx experimental configurations respectively. The dashed horizontal magenta line shows the emission from LAE galaxies \citep[e.g.,][]{heneka2021} around this redshift { ($z \sim 10$)}.}
   \label{fig:autohilae1}
\end{figure}

The pixel volume is defined in terms of the parameters of the instrument,  as
\citep[e.g.,][]{dumitru2019}:
\begin{eqnarray}
V_{\rm pix} &=& 1.1 \times 10^3  {\rm (cMpc}/h)^3 \left (\frac{\lambda}{158  \ 
\mu m} \right) \left(\frac{1 +z}{8} \right)^{1/2}  \nonumber \\
&& \left(\frac{\theta_{\rm beam}}{10 '} \right)^2 \left(\frac{\delta \nu}{400 
{\rm MHz}} \right) \, ,
\label{eqnvpix}
\end{eqnarray}
in which $\lambda$ is the rest wavelength of the line (here Lyman-$\alpha$) and $\delta \nu$ is the spectral width of the observation. The $\theta_{\rm beam}$ is the beam size, given by the observed wavelength $\lambda_{\rm obs}$ and the diameter of the telescope dish, as $\theta_{\rm beam} = 1.22 \lambda_{\rm obs}/D_{\rm dish}$.

{ The noise parameters of the various surveys under consideration are listed in Table \ref{table:improved} and briefly described here. } For the CDIM-like survey, we use the parameters of the proposed Cosmic Dawn Intensity Mapper \citep[CDIM;][]{cdimreport2019} which covers the bandwidth $0.75 -7.5 \mu$m with 840 spectral channels and surveys a 300 deg$^2$ area of the sky.  
{ SPHEREx\footnote{https://spherex.caltech.edu/page/instrument} is described following the specifications for the all-sky survey in \citet{cheng2022}.
It assumes a noise sensitivity of 10 nW m$^2$/sr in the wavelength range centred around $2.5 \ \mu$m,  which translates to $\sigma_{\rm N} = 8.33 \times 10^{-20}$ in units of erg cm$^{-2}$ s$^{-1}$ Hz$^{-1}$ / sr.
It also assumes  spectral channels of $\Delta \lambda = 0.04 \ \mu$m around the central wavelength, with a bandwidth covering the 1-5  $\mu$m range.}

The power spectrum is modulated by the effects of the finite beam of the experiment \citep[see also, e.g.,][]{hpoiii, hp2023im} \footnote{ The present analysis of the beam effects is equivalent to other treatments for the window function used in the intensity mapping literature, e.g., \citet{lidz2011, heneka2017}.  For a detailed explanation of this comparison, see e.g., Appendix C.3 of \citet{li2015}.}:
 \begin{eqnarray}
 W_{\rm beam}(k) &=& e^{-k^2 \sigma_{\perp}^2} \int_0^1 e^{-k^2 \mu^2 (\sigma_{\parallel}^2 - \sigma_{\perp}^2)} d \mu \nonumber \\
   \noindent &=& \frac{1}{k \sqrt{\sigma_{\parallel}^2 - \sigma_{\perp}^2}} \frac{\sqrt{\pi}}{2} {\rm{Erf}} \left(k  \sqrt{\sigma_{\parallel}^2 - \sigma_{\perp}^2} \right)  \nonumber \\
   &\times& \exp(-k^2 \sigma_{\perp}^2) \, .
   \label{wij}
\end{eqnarray}
In the above, the terms $\sigma_{\perp}$ and $\sigma_{\parallel}$ account for the finite spatial and spectral resolution \citep{li2015}:
\begin{equation}
    \sigma_{\perp} = \chi(z) \sigma_{\rm beam} \, ,
    \label{sigmaperp}
\end{equation}
with $\chi(z)$ being the comoving distance to redshift $z$, $\sigma_{\rm beam} = \theta_{\rm beam}/\sqrt{8 \ln 2}$, and
\begin{equation}
    \sigma_{\parallel} = \frac{c}{H(z)} \frac{(1+z)^2 \delta \nu}{\nu_{\rm obs}} \, .
    \label{sigmapar}
\end{equation}
The other correction factor which accounts for the finite volume of the survey is given by ({ see, e.g., \citet{bernal2019} for a description of the expression and its use in the intensity mapping literature}):
\begin{equation}
    \begin{split}
W_{\rm vol}(k,\mu) = & \left(1-\exp\left\lbrace -\left(\frac{k}{k^{\rm min}_\perp}\right)^2\left(1-\mu^2 \right) \right\rbrace \right)\times \\
& \times \left(1-\exp\left\lbrace -\left(\frac{k}{k^{\rm min}_\parallel}\right)^2\mu^2 \right\rbrace \right).
\end{split}
\label{eq:Wk_vol}
\end{equation}
{ Here,  $k^{\rm min}_{\perp} \equiv 2\pi/L_{\perp}$,  and $k^{\rm min}_{\parallel} \equiv 2\pi/L_{\parallel}$, with $L_\perp$ and $L_\parallel$ being the maximum transverse and radial length scales probed by the survey in the spatial and frequency directions, respectively.}
The volume of the survey is then approximately given by $L_\perp^2 L_\parallel$. The lengthscales $L_\perp$ and $L_\parallel$  are defined by:  
\begin{equation}
    L_{\parallel} = \frac{c}{H(z)} \frac{(1+z) B_{\nu}}{\nu_{\rm obs}} \, ,
\end{equation}
and 
\begin{equation}
    L_{\perp}^2 = \chi^2(z) \Omega_{A} \, ,
\end{equation}
where $B_{\nu}$ is the bandwidth of the survey and
$\Omega_{A}$ is the solid angle associated with the survey area,  $\Omega_{A} = S_{A}/4\pi$. This leads to the expression (averaged over the angular variable):
\begin{eqnarray}
   && W_{\rm vol}(k) = 1 - \exp(-k^2/k_{\perp}^2)  \int_0^1 d \mu \exp (k^2 \mu^2 /k_{\perp})^2 \nonumber \\
   &+& \exp(-k^2/k_{\perp}^2) \int_0^1 \exp - \left(k^2 \mu^2 /k_{\parallel}^2 - k^2 \mu^2/k_{\perp}^2\right) d \mu \nonumber \\
  & \approx &  1 - \frac{\sqrt{ \pi} k_{\parallel}}{2 k}{\rm{Erf}} \left(k / k_{\parallel} \right)
\end{eqnarray}
{ where we have suppressed the ``min" subscripts on $k_{\parallel}$ and $k_{\perp}$ for simplicity and the second line is valid for $k_{\parallel} >> k_{\perp}$, representative of surveys having a large area on the sky.  
}

The full window function associated with the spatial and spectral effects of the instrument is then given by:
\begin{equation}
W_{\rm Lya}(k) = W_{\rm beam} (k) W_{\rm vol}(k) \, .
\label{fullwindowfunction}
\end{equation}
{ With all the components in place,} the overall variance in the autocorrelation  power spectrum is given by \citep[e.g.,][]{lidz2011, gong2012, dumitru2018}:
\begin{equation}
{\rm var}_{\rm LR, Lya}  = (P_{\rm LR, Lya}W_{\rm Lya}(k) + P^{\rm Lya}_{\rm N})^2/N_{\rm modes} \, ,
\label{varianceauto}
\end{equation}
with the number of modes, $N_{\rm modes}$ defined as:
\begin{equation}
N_{\rm modes} = 2 \pi k^2 \Delta k V_{\rm surv}/(2 \pi)^3 \, .
\end{equation}
In writing the above, we have assumed the $k$ values to be logarithmically equispaced with an interval of $\Delta \log_{10} k = 0.2$,  with
the survey volume, $V_{\rm surv}$ defined by:
\begin{eqnarray}
V_{\rm surv} &=& 3.7 \times 10^7  {\rm (cMpc}/h)^3 \left (\frac{\lambda}{158  \ 
\mu {\rm m}} \right) \left(\frac{1 +z}{8} \right)^{1/2}  \nonumber \\
&& \left(\frac{S_{\rm{A}}}{16 {\rm deg}^2} \right) \left(\frac{B_{\nu}}{20 
{\rm GHz}} \right) \, ,
\label{vsurveycross}
\end{eqnarray}
in terms of the instrumental parameters specified earlier.

\begin{figure}
    \centering
    \includegraphics[width =\columnwidth]{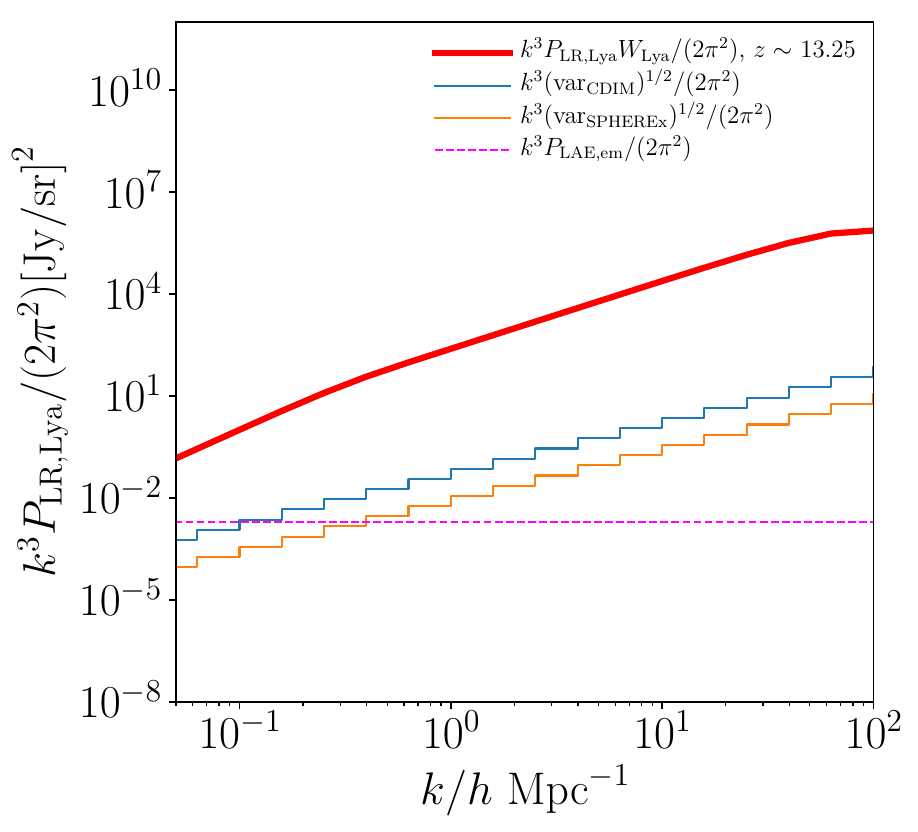}   \includegraphics[width = \columnwidth]{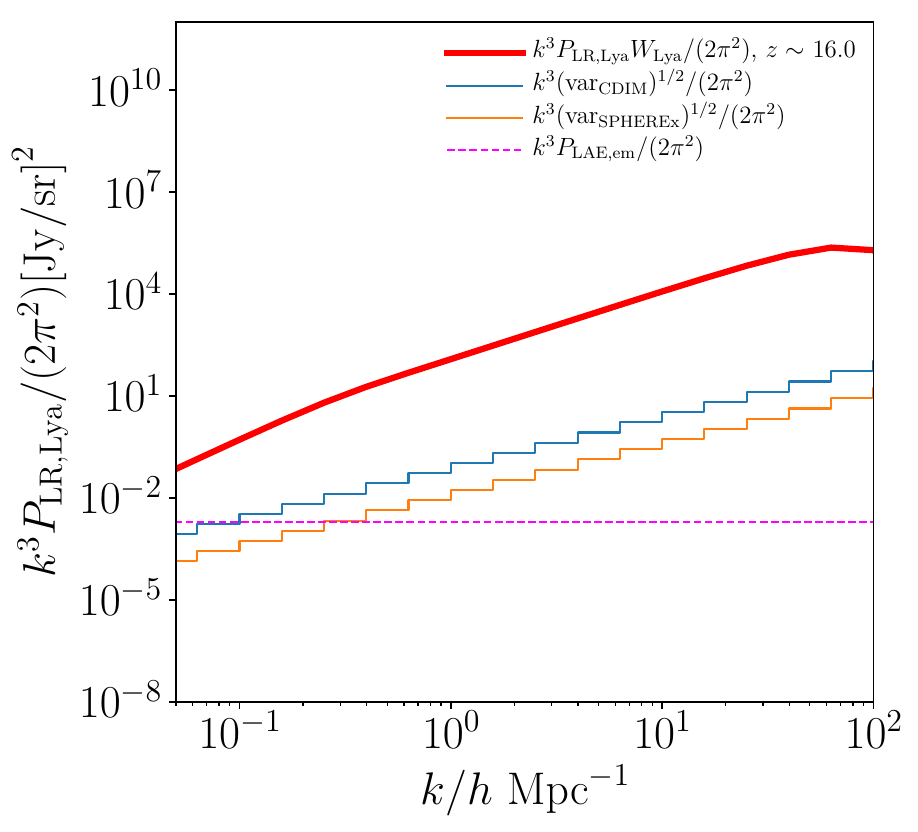} 
    \caption{ Same as Fig. \ref{fig:autohilae1}, for Lyman-$\alpha$ haloes at $z \sim 13$ (top panel) and $z \sim 16$ (lower panel) respectively.}
   \label{fig:autohilae2}
\end{figure}

{ Figures \ref{fig:autohilae1} and \ref{fig:autohilae2} show the intensity mapping power spectra, \eq{powerspeclr}  at each of the redshifts $z \sim 9, 10.5, 13$ and 16 { modulated by a large survey area (CDIM-like/SPHEREx-like) window function (thick solid red lines)}.  In each figure, the thinner blue and orange steps show the square root of the noise variance, given by \eq{varianceauto}, and corresponding to the CDIM-like and SPHEREx experimental configurations. } { Overplotted for comparison is the intensity mapping power spectrum due to Lyman-alpha emitters alone (the LAE signal) at these redshifts \citep{heneka2021}.  At $z \gtrsim 10$,  the LAE signal is found to be $\Delta_{\rm LAE,em} \leq 10^{-10}$ erg s$^{-1}$ cm$^{-2}$ sr$^{-1}$,  roughly constant over the whole $k$-range [see e.g., Fig. 2, middle right panel of \citet{heneka2021}].  The power spectrum is the square of this quantity, which, in order to bring the units in line with those in the present work,  needs to be divided by the square of the observed Lyman-$\alpha$ frequency,  viz. $\nu_{\rm obs}^2$ (Caroline Heneka, private communication). This leads to $k^3 P_{\rm LAE,em}/(2 \pi^2) \sim 2 \times 10^{-3}$ (Jy/sr)$^2$ in the present units, which is plotted as the dashed horizontal magenta line in Figures \ref{fig:autohilae1} and \ref{fig:autohilae2}.  It can be seen that the  intergalactic signal from the Loeb-Rybicki haloes is expected to dominate at all the redshifts and scales under consideration.}

{{\subsection{Polarization signal from Loeb-Rybicki haloes}

\begin{figure}
    \centering
    \includegraphics[width =\columnwidth]{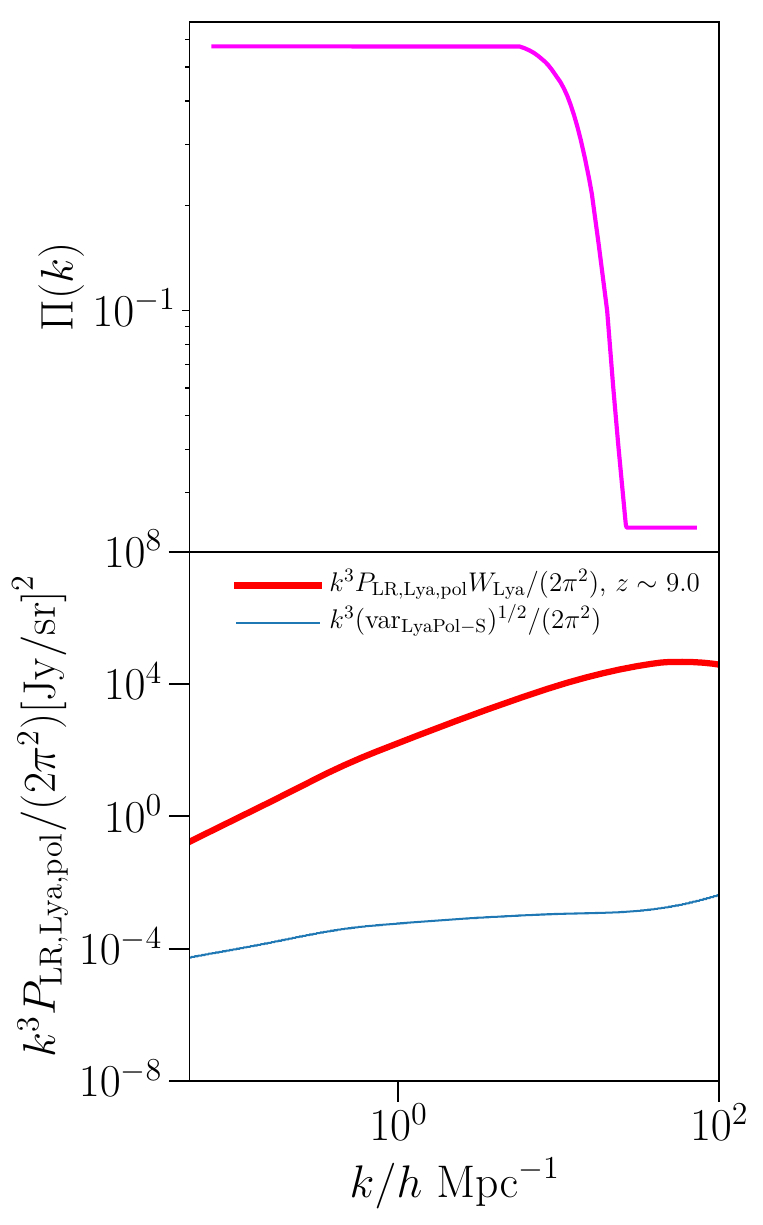}    
    \caption{ {\textit{Top panel:} Polarization fraction $\Pi(k)$ associated with Loeb-Rybicki haloes as a function of (inverse) scale $k$. \textit{Lower panel:} Integrated intensity of the polarized component of the Loeb-Rybicki signal (red solid curve) and the square root of the noise variance associated with a hypothetical spaced based experiment, LyaPol-S, whose parameters are described in Table \ref{table:polarization}.}}
   \label{fig:piofk}
\end{figure}

The Loeb-Rybicki haloes are expected to be highly polarized \citep{rybicki1999}, which could add another path towards improving the signal-to-noise ratio for their detection.  Here, we explore some aspects of this possibility.

\begin{table*}
\begin{center}
    \begin{tabular}{ | c | c | c | c | c | c | c | c | c |  p{5 cm} |}
    \hline
    Configuration &  $D_{\rm dish}$ (m.) & $\delta \nu$ (GHz)  & $S_{\rm A}$ (sq. deg.) & $\sigma_{\rm N}$ 
(erg cm$^{-2}$ s$^{-1}$ Hz$^{-1}$ / sr) & $B_{\nu}$   \\ \hline
LyaPol-S &  0.20 & 400 &  330 & $6.5 \times 10^{-22}$  & 200000 GHz  \\ 
\hline
    \end{tabular}
\end{center}
\caption{ {Experimental parameters assumed for a hypothetical space-based polarization survey design, termed  LyaPol-S in \citet{masribas2020}.}}
 \label{table:polarization}
\end{table*}

The experiments considered thus far are not designed to measure polarization.  However, we can make an estimate of the signal-to-noise achievable by a hypothetical space-based experiment (with the specifications of the so-called Lyapol-S, \citet{masribas2020}, summarized in Table \ref{table:polarization}) designed to probe the polarized component of intensity of Loeb-Rybicki haloes at $z \sim 9$.  

Using a Monte Carlo analysis of polarized radiative transfer, it is possible to obtain the profile of the (frequency-integrated) polarized intensity of the Loeb-Rybicki halo \citep{rybicki1999}, as well as the degree of polarization $\Pi(\tilde{p})$ as a function of the (dimensionless) impact parameter, $\tilde{p}$.  This enables the polarized intensity to be expressed as a function of the (inverse) scale $k$,  where $k = 2 \pi/r$.  The evolution of $\Pi(k)$, which denotes the (fractional) degree of polarization of the radiation expected as a function of distance from the halo, is displayed in the top panel of Fig. \ref{fig:piofk} for $z \sim 9$. 

By  multiplying the intensity of the radiation, $I_0$ in \eq{powerspeclr}  by the term $\Pi(k)$,  we get the intensity of polarization as a function of impact parameter  \citep{masribas2020}.
The power spectrum of the polarization intensity fluctuations, denoted by $P_{\rm LR, Lya, pol} (k)$ can, in turn,  be obtained from the intensity of polarization by following the same procedure as that leading to \eq{powerspeclr}.  This is plotted for $z \sim 9$ as the thick red solid line in the lower panel of Fig. \ref{fig:piofk}.  Also plotted as the thin blue line on the same panel is the  the square root of the noise variance associated with the LyaPol-S experiment,  denoted by (var$_{\rm LyaPol-S}$)$^{1/2}$ and obtained by following the same procedure as that leading to \eq{varianceauto}.  The noise term for this case is calculated by using  \eq{noiselyaauto} with the specifications of LyaPol-S (given in Table \ref{table:polarization}). The plot confirms the high signal-to-noise of  detectability of the polarization of the Loeb-Rybicki haloes with the hypothetical experiment constructed above.}
}}

\subsection{Cross-correlation with 21 cm intensity mapping}

So far, we have seen that there are good prospects for detecting the autocorrelation signal from the intergalactic Lyman-$\alpha$ haloes in intensity mapping out to $z \sim 16$ with current and near-future facilities.  We now examine the prospects for cross-correlating this signal with the 21-cm radiation coming from the diffuse HI in the IGM around these haloes.

The neutral hydrogen at the epoch of reionization mostly resides in the diffuse IGM.  If we assume that the HI follows dark matter, i.e. $b_{\rm HI} = 1$ across all the redshifts under consideration,  the power spectrum of its 21-cm emission can be expressed as:
\begin{equation}
P_{\rm HI, IGM} = P_{\rm dm}(k,z)  T_0^2(z) \, ,
\label{phiigm}
\end{equation}
where \citep[e.g.,][]{pritchard2012}
\begin{equation}
T_0(z) = 27 \  {\rm mK} \ \frac{\Omega_b h^2}{0.023} \  \frac{0.15}{\Omega_m h^2}\left(\frac{1+z}{10}\right)^{0.5} \, .
\label{t0hi}
\end{equation}
The temperature normalization of \eq{phiigm} may be converted from units of mK$^2$ to (Jy/sr)$^2$ by converting the $T^2_\mathrm{0}(z)$ term into intensity via
$I_{\rm 0,  21 cm} = {2 k_B T_{\rm 0}(z)}/{\lambda(z)^2}$,
with  $\lambda(z) \equiv 21 {\rm cm}(1+z)$.  
{ \eq{t0hi} assumes that the spin temperature of the HI gas (denoted by $T_S$)  at these epochs is larger than the background CMB temperature (denoted generically by $T_R$). This requires the IGM to be  X-ray heated  (as expected at $z \lesssim 20$). Relaxing the heating assumption leads to the introduction of a $(1 - T_R/T_S)$ term on the RHS of \eq{t0hi}. While observational constraints on the spin temperature are loose at the moment \citep{hera2022},  we can use the latest data from the Hydrogen Epoch of Reionization Array (HERA) to infer the maximum effect of the $(1 - T_R/T_S)$ term at $z \sim 10.5$ as described below.}

The cross-correlation signal between the Loeb-Rybicki haloes and the 21 cm emission from HI in the IGM at a given redshift can now be expressed as:
\begin{equation}
    P_{\times} = (P_{\rm RL, HI-Lya} P_{\rm HI, IGM})^{1/2} \, .
    \label{crosscorrpow}
\end{equation}

The cross-correlation signal expression is equivalent, modulo a shot-noise like contribution, to that obtained by  averaging the contributions independently before multiplication - for more details, see e.g.,  \citet{liu2021, beane2018, hpoiii, oxholm2021}. The additional contributory term is, in turn, expected to be subdominant on the clustering scales \citep[e.g.,][]{liu2021} from where the bulk of the cross-correlation signal originates.  {  We follow \citet{breysse2022} and \citet{hpoiii} in assuming that both intensity maps can be treated as Gaussian for the purpose of estimating the binned power spectrum at the scales of interest. While individual Fourier voxels are clearly non-Gaussian, binning over many modes produces an approximately Gaussian distribution near the peak, as demonstrated in \citet{fronenberg2024}, with deviations mainly in the tails. This approximation is expected to hold on large, linear or weakly non-linear scales and at early times, when the 21\,cm signal more closely traces the underlying matter field and astrophysical effects can be captured by an effective bias, consistently with our present treatment. 

}

We now investigate the detectability of the cross-power spectrum with current and upcoming facilities.  We consider a Murchison Widefield Array (MWA)-like survey, as well as its successor, the Square Kilometre Array (SKA)-LOW, each cross-correlated with the three Lyman-$\alpha$ intensity mapping surveys discussed in the previous section.

\begin{figure}
    \centering
     \includegraphics[width =\columnwidth]{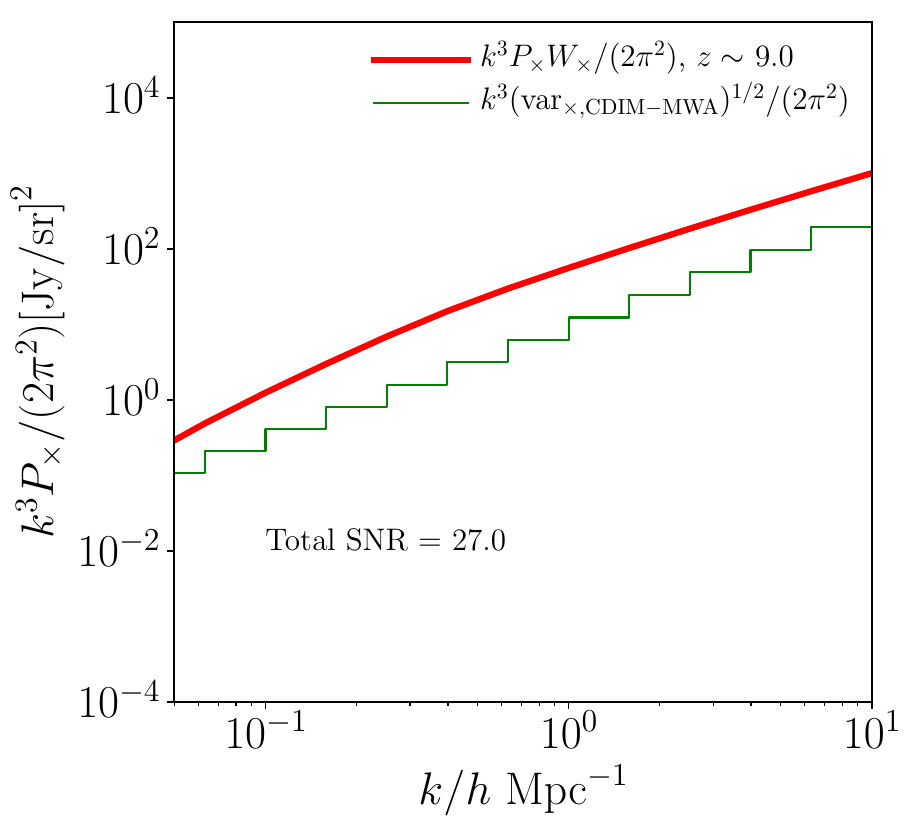}  
     \includegraphics[width = \columnwidth]{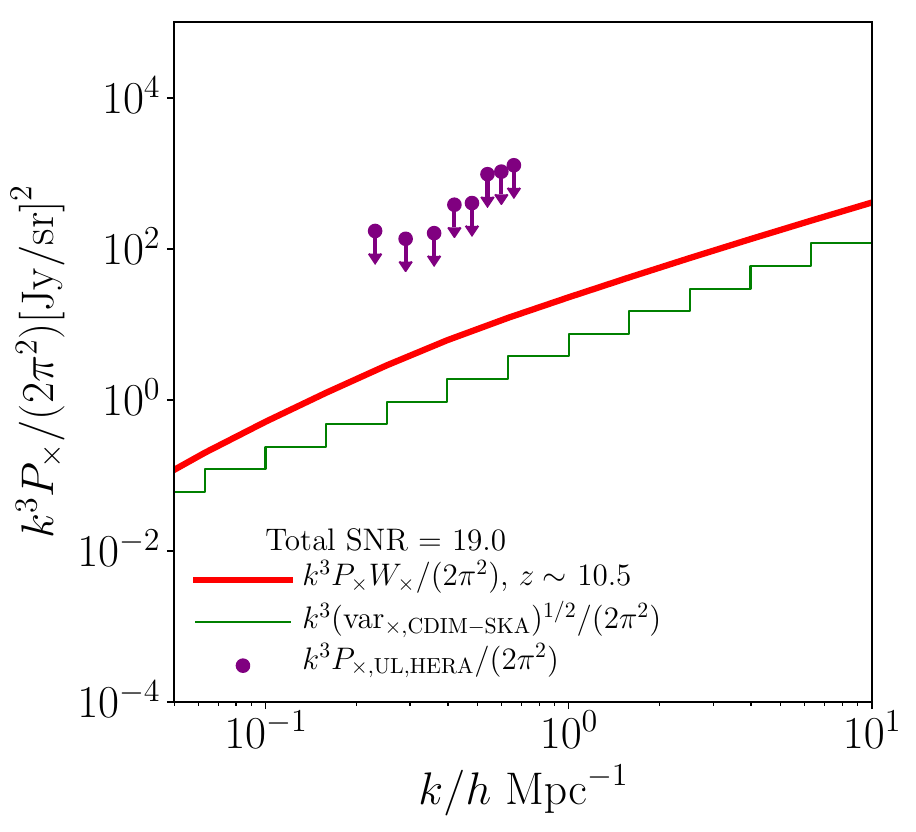} 
    \caption{Cross correlation signal (thick red lines) and noise (thin green lines) variance between 21 cm and Lyman-$\alpha$ haloes at $z \sim 9$ probed by a CDIM-like-MWA experiment combination (top panel) and at $z \sim 10.5$ probed by a CDIM-like-SKA experiment combination (lower panel). { The lower panel also shows the upper limits on the cross-correlation power spectrum (purple error bars)  arising from the corresponding ones on the 21-cm signal at $z \sim 10.4$ from the latest results of the \citet{hera2023}, which represent the (maximum) effect of including the spin  temperature term, $(1- T_R/T_S)$ in \eq{t0hi}. }}
     \label{fig:21cmlyacc1}
\end{figure}

\begin{table*}
\centering
\begin{tabular}{cccccccccc}
\hline
Configuration & $d_{\rm max}$ & $N_{\rm a}$ & $n_{\rm pol}$ & $T_{\rm inst}$[K]  & $A_{\rm eff}$ (m$^2$) & $t_{\rm obs}$[h] &     $S_{\rm A}$[deg.sq] \\ \hline
MWA & 1000 m & 256 &  2 & 28 &  14.5 & 2000 & 1000  \\
SKA-LOW &  40000 m &  512  &  2 & 28  & 964 & 2000  & 1000 \\
\hline
\end{tabular}
\caption{Noise parameters for the 21 cm surveys, following \citet{hp2023im}. $N_{\rm a}$ here denotes the number of independent elements,  viz.  antennas for the MWA and  SKA-LOW \citep[e.g.,][]{furlanetto2007}.}
\label{table:hi}
\end{table*}

 The thermal noise of an  interferometric 21 cm survey is given by \citep[e.g.,][]{bull2014, obuljen2018}:
\begin{equation}
P^{\rm{HI}}_{\rm{N}}(z) = \frac{T^2_\mathrm{sys}(z)\chi^2(z)r_\nu(z) \lambda^4(z)}{A^2_{\rm{eff}} t_{\rm obs} n_{\rm{pol}}n(\textbf{u},z) \nu_{\rm 21}},
\label{noisehiauto}
\end{equation}

In the above expression,  $A_{\rm eff}$ denotes the effective collecting area of a single (antenna) element, with  $T_{\rm sys}= T_{\rm sky}+T_{\rm inst}$ being the system temperature of the receiver.  In the above, $T_{\rm sky} = 60{\rm K} \big(300 {\rm MHz}/\nu\big)^{2.25}$ is the sky contribution to the system temperature and $\nu = 1420 {\rm MHz}/(1+z) \equiv \nu_{21}/(1+z)$ is the observed frequency, corresponding to the observed wavelength $\lambda (z) \equiv 21 \ {\rm cm}  (1+z)$.  The instrumental parameters for the MWA and SKA-LOW surveys under consideration are provided in Table \ref{table:hi}. 
For the MWA, we adopt the Phase II configuration \citep{mwa2018, mwa2019} and for the SKA-LOW,  a configuration of 512 antennas with effective area $962$ m$^2$ each,  based on SKA I Level 0 specifications\footnote{https://www.skao.int/sites/default/files/documents/d4-SKA-TEL-SKO-0000007\_SKA1\_Level\_0\_Science\_RequirementsRev02-part-1-signed\_0.pdf} (see \citet{hp2023im} for a detailed description of these configurations).  Both surveys are assumed to run for $t_{\rm obs} =  2000$ h, roughly corresponding to two years observing time.

In \eq{noisehiauto}, $r_{\nu}(z)$ is a redshift-dependent factor that converts from bandwidth to survey depth:
\begin{equation}
r_{\nu}(z) = \frac{c (1 + z)^2}{H(z)} \, ,
\end{equation}
with $H(z)$ being the Hubble parameter at redshift $z$.  The term $n({\textbf{u}}, z)$ denotes the baseline density in visibility space,  normalized to the number of independent elements $N_{\rm a}$ and approximated by:
\begin{equation}
n(u, z) = \frac{N_{\rm a}^2}{2 \pi u_{\rm max}^2} \, ,
\end{equation}
with $u_{\rm max}$  being related to the maximum baseline, $d_{\rm max}$, by
\begin{equation}
u_{\rm max} = \lambda (z)/d_{\rm max} \, .
\end{equation}
The number of independent polarizations, $n_{\rm pol}$,  is set to 2 for both surveys we consider here.
The noise power in \eq{noisehiauto} may be converted to units of (Jy/sr)$^2$  by converting the $T^2_\mathrm{sys}$ term into intensity { in a similar manner as done the signal (see discussion following \eq{t0hi})}.

For the cross-correlation measurement,  we define the number of modes by using the overlapping survey volume $V_{\rm surv, \times}$, which is taken to be the volume associated with the smaller survey:
\begin{equation}
N_{\rm modes, \times} = 2 \pi k^2 \Delta k \frac{V_{\rm surv, \times}}{(2 \pi)^3} \, ,
\label{nmodes}
\end{equation}
{In our sensitivity forecasts we may account for contamination of cosmological Fourier modes by spectrally smooth foreground emission.  To model the loss of line-of-sight modes that are intrinsically contaminated by foregrounds we excise modes with
\[
k_\parallel\;\le\;0.07\ \mathrm{Mpc}^{-1}
\]
in the fiducial and optimistic foreground scenarios and
\[
k_\parallel\;\le\;0.175\ \mathrm{Mpc}^{-1}
\]
in the pessimistic scenario, following the intrinsic foreground treatment described in \citet{Kubota2018}. These cuts reduce the effective number of accessible Fourier modes in our variance and signal–to–noise calculations,  becoming relevant when the volume of the HI survey is the dominant one in cross-correlation.
}
This modifies the ``effective" volume of the survey to:
\begin{eqnarray}
V_{\rm surv, \times} &=& 3.7 \times 10^7 f_{\mu} {\rm (cMpc}/h)^3 \left (\frac{\lambda_{\times}}{158  \ 
\mu {\rm m}} \right) \left(\frac{1 +z}{8} \right)^{1/2}  \nonumber \\
&& \left(\frac{S_{\rm{A, \times}}}{16 {\rm deg}^2} \right) \left(\frac{B_{\nu, \times}}{20 
{\rm GHz}} \right) \, ,
\label{vsurveycross}
\end{eqnarray}
with the parameters of the smaller volume survey being expressed using the $\times$ subscript, as $S_{A, \times}, \lambda_{\times}, $ and $B_{\nu, \times}$ and { $f_{\mu}$ being a foreground sensitive parameter, invoked when the HI survey is the smaller volume survey,  and set to unity otherwise}.
Accounting  for the finite resolution and volume effects given by \eq{fullwindowfunction}, the variance of the cross-correlation is given by \citep[e.g.,][]{hpoiii}:
\begin{eqnarray}
    {\rm var}_{\times}(k) &=& \left((P_{\rm RL, HI-Lya}W_{\rm Lya}(k) + P_{\rm N}^{\rm Lya}) \right. \nonumber\\
   & & \left. (P_{\rm HI, IGM} + P_{\rm N}^{\rm HI}) \right. \nonumber\\
    &+& \left. P_{\times}^2 W_{\rm Lya}(k) \right)/2 N_{\rm modes, \times} \, ,
\label{varnoisecross}
\end{eqnarray}
in which $W_{\rm Lya}(k),  P^{\rm N}_{\rm HI}$  and $P_{\rm N}^{\rm Lya}$ follow \eq{fullwindowfunction}, \eq{noisehiauto} and \eq{noiselyaauto} respectively. Given the above signal and noise variance, the signal-to-noise ratio (SNR) of the cross-correlation measurement is calculated as:
\begin{equation}
{\rm{SNR}} = \left(\sum_k \frac{P_{\times}^2(k) W_{\times}^2(k)}{{\rm{var}}_{\times}(k)}\right)^{1/2} \, ,
    \label{snrcross}
\end{equation}
in which the window function $W_{\times}^2(k) \equiv W_{\rm Lya}(k)$ modulates the cross-power spectrum. {\footnote{ For the $k$-range under consideration and the instrument parameters,  $W_{\rm HI} = 1$ is found to be a good approximation, see also \citet{hp2023im}.}}

\begin{figure}
    \centering
    \includegraphics[width =\columnwidth]{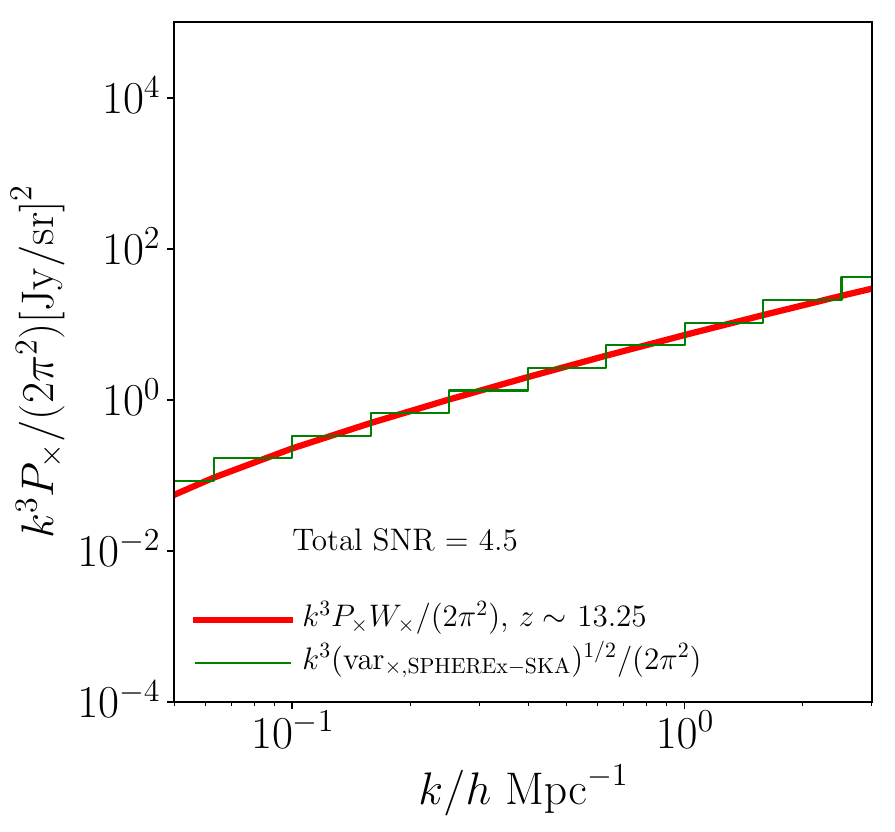}   \includegraphics[width = \columnwidth]{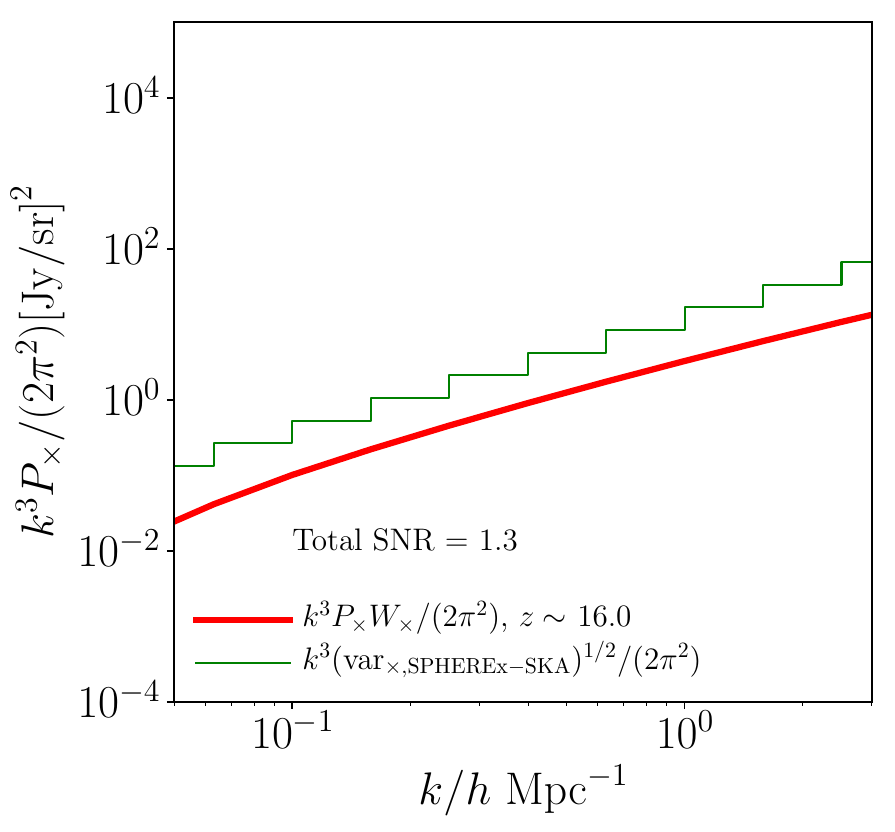} 
    \caption{Cross correlation signal (thick red lines) and noise (thin green lines) variance between 21 cm and Lyman-$\alpha$ haloes at $z \sim 13$ (top panel) and 16 (lower panel),  probed by a SPHEREx-SKA experiment combination.}
    \label{fig:21cmlyacc2}
\end{figure}

The cross-correlation power spectra modulated by the window function, $P_{\times} W_{\times}$ are plotted in Figs. \ref{fig:21cmlyacc1}, \ref{fig:21cmlyacc2} at $z \sim 9, 10.5, 13.25$ and 16 respectively as the red solid lines.  The thinner green lines show the square root of the cross-variance of the noise, given by \eq{varnoisecross} for different experiment combinations (CDIM-MWA,  CDIM-SKA and SPHEREx-SKA).  Listed in each figure is the SNR of the corresponding configuration. 
{ At $z \sim 10.4$,  the HERA results find 95\% confidence upper limits on the 21 cm power spectrum of $\Delta^2$ ($k = 0.36 h$ Mpc$^{-1}$) < 3496 mK$^2$.  This can be used to provide an estimate of the (maximum) effect of the $(1 - T_R/T_S)$ term which was set to unity in \eq{t0hi}.  By computing the cross-correlation of the measured 21 cm upper limits and the Lyman-$\alpha$ intensity mapping power spectrum at this redshift, we obtain the purple upper limits shown in the lower panel of Fig. \ref{fig:21cmlyacc1}.  These lie about 1-2 orders of magnitude above the fiducial power spectrum (red curve in the same figure). Our findings are consistent with the 3 orders of magnitude difference between the HERA upper limits and fiducial 21 cm models \citep{hera2023}, since the square root of the 21 cm power appears in the expression for the cross-power spectrum.}

It is seen that the above experiment combinations lead to a few { to a few} tens of standard deviations detection of the cross-power spectrum out to $z \sim 13$,  { with marginal detectability out to $z \sim 16$.} The detectability of the cross-power is  important to secure an unambiguous confirmation of the 21 cm signal at these redshifts,  especially given the challenges posed by foregrounds to the radio intensity mapping surveys.

\section{Discussion}
\label{sec:discussion}

In this paper, we have explored the prospects for detecting the { integrated Lyman-$\alpha$ signal from the clustering of  intergalactic Loeb-Rybicki haloes}  via intensity mapping before and during the epoch of reionization ($z \sim 9-16$).  We have found that current and future experiments probing Lyman-$\alpha$ intensity (such as the SPHEREx and CDIM-like experiments) offer excellent prospects for the detectability of this signal, both in autocorrelation as well as cross-correlations with upcoming 21-cm facilities such as the Murchison Widefield Array (MWA) and the Square Kilometre Array (SKA)-LOW. 
The intensity mapping power spectrum from the intergalactic  haloes  is stronger than the emission coming from the LAE galaxies themselves, the latter of which is expected to be sharply attenuated due to the increasing neutral hydrogen fraction in the IGM before reionization.  Probing the `break' scale between the one- and two-halo terms in the power spectrum (Fig.  \ref{fig:fidpower}) also offers the possibility to constrain the typical size of the Loeb-Rybicki haloes.

Mapping the intensity of intergalactic haloes (either in autocorrelation or cross-correlation with the 21-cm radiation emitted by the neutral hydrogen itself) is thus expected to serve as a useful new diagnostic of the IGM prior to reionization.  { The cross-correlation is especially important considering that the 21-cm foregrounds, which present a challenge for avoidance or removal at these low frequencies \citep{liu2020}, largely originate from low redshifts, hence reionization-era surveys have very low probabilities of having shared foregrounds.  Hence,  we assume that observational systematics involving foregrounds are mitigated in the cross-correlation survey.  Nevertheless,  the foregrounds can affect the variance of the cross-power spectrum, which requires a more detailed analysis than the one described here,  and involves the product of the Lyman-alpha interloper or foreground radiation with the 21 cm signal.  At these wavelengths,  the major sources of foreground radiation are the zodiacal light in the Solar system and the diffuse galactic light arising from dust scattering in the Milky Way \citep{masribas2020, rigby2023}, both of which are possible to model and subtract due to their relatively smooth spectrum with known frequency dependences \citep{brandt2012}.  Bright sources contributing to the interloper radiation is also expected to be easy to identify and mask in order to avoid contamination \citep{masribas2020}, similarly to the case of sub-millimetre tracers \citep[see, e.g.,][]{sun2018}, with a minimal drop in the survey volume.} The cross-correlation measurement can help in an unambiguous confirmation of the as-yet elusive 21-cm signal, in a cosmological regime where other tracers of interest for intensity mapping cross-correlations (such as the fine-structure lines of [OIII], [CII] and the rotational transitions of CO) are not present in significant amounts.  

{The assumed sensitivity for the CDIM-like survey adopted in our fiducial calculations reflects conservative survey choices rather than the instrument’s intrinsic capability.  In practice, CDIM’s larger mirror, higher spectral resolution, and smaller beam could achieve better sensitivity than SPHEREx over the same redshift range.  Sensitivity also varies modestly with redshift due to its dependence on the shifting observed wavelength.  Even so, the fiducial assumptions are sufficient for cross-correlation studies with 21\,cm surveys, and optimized CDIM observations would only improve these constraints.}

In this paper, we have followed a similar approach to \citet{hp2023im} in focusing on the magnitude of the 21-cm signal rather than the details of its scale dependence, which are at present poorly constrained.  Accounting for { the absence of neutral hydrogen in ionized bubbles would 
be expected to boost the cross-correlation signal} by factors of up to $\sim 30$ on scales $k \sim 0.1$ Mpc/$h$, over and above the estimates here \citep{lidz2008, hp2023im}.  A similar boost in the signal-to-noise may occur from an increase in the sky areal coverage of the MWA and SKA-LOW surveys to $\sim 25000$ deg$^2$ (from the currently assumed $\sim 1000$ deg$^2$). { In a more detailed analysis,  the region of overlapping coverage between the 21 cm and Lyman-$\alpha$ surveys needs to be taken carefully into consideration, instead of the full overlap implicitly assumed in the present work.  We have adopted the isotropic mode-counting and window-function approximations commonly used in the literature \citep[e.g.,][]{li2015}. The surveys that determine our main  result, viz. CDIM-like, MWA/SKA, and SPHEREx, are all wide-field experiments, for which the dominant signal-to-noise contribution comes from scales where these approximations are valid. In this regime, anisotropies in cylindrical power spectra and survey window functions produce only subdominant corrections and do not affect our conclusions. { While the mode count in our present cross-correlation analysis is limited by the LAE survey size, we note that is not generically expected to be the case.  In particular,  including a more realistic wedge treatment of the 21 cm foreground contamination [following,  e.g.,  \citet{Pober2014, Liu2014a, Pober2015}]  would further reduce the available mode volume and potentially impact the forecasted signal-to-noise ratios.} 
} We have also assumed that peculiar velocities of the hydrogen gas may be neglected in \eq{t0hi}.  Addressing these effects, as well as accounting for the evolution of ionized bubbles in the IGM  which lead to a turnover in the power spectrum on small-scales \citep{lidz2011, dumitru2019}, will be the subject of future work.

\section*{Acknowledgements}

We thank Dongwoo Chung and Caroline Heneka for helpful clarifications regarding experimental configurations, { and the referee for a detailed and helpful report that improved the content and presentation of the paper}.  HP's research was supported by the Swiss National Science Foundation via Ambizione Grant PZ00P2\_179934. The work of AL was partially supported by the Black Hole Initiative at Harvard 
University, which is funded by grants from the JTF and GBMF.

\def\aj{AJ}                   
\def\araa{ARA\&A}             
\def\apj{ApJ}                 
\def\apjl{ApJ}                
\def\apjs{ApJS}               
\def\ao{Appl.Optics}          
\def\apss{Ap\&SS}             
\def\aap{A\&A}                
\def\aapr{A\&A~Rev.}          
\def\aaps{A\&AS}              
\def\azh{AZh}                 
\def\baas{BAAS}
\def\jcap{JCAP}
\def\jrasc{JRASC}             
\def\memras{MmRAS}
\def\na{New Astronomy}
\def\nat{Nature}
\def\mnras{MNRAS}             
\def\pra{Phys.Rev.A}          
\def\prb{Phys.Rev.B}          
\def\prc{Phys.Rev.C}          
\def\prd{Phys.Rev.D}          
\def\prl{Phys.Rev.Lett}       
\def\pasp{PASP}               
\def\pasj{PASJ}
\def\physrep{Phys. Repts.}
\def\qjras{QJRAS}             
\def\skytel{S\&T}             
\def\solphys{Solar~Phys.}     
\def\sovast{Soviet~Ast.}      
\def\ssr{Space~Sci.Rev.}      
\def\zap{ZAp}                 
\let\astap=\aap
\let\apjlett=\apjl
\let\apjsupp=\apjs

\small{
\bibliographystyle{aa}
\bibliography{mybib, main, refs, biblio, mybib2}{}
}

\label{lastpage}
\end{document}